\documentclass[11pt, aps, prfluids, amsmath, amssymb, superscriptaddress, onecolumn,floatfix,tightenlines,showpacs,longbibliography,notitlepage]{revtex4-2}

\usepackage{graphicx}
\usepackage{dcolumn}
\usepackage{bm}
\usepackage{soul,placeins}
\usepackage{xcolor}
\usepackage{siunitx}
\usepackage[breaklinks,colorlinks = true,linkcolor = blue,urlcolor=blue,citecolor=blue]{hyperref}

\newcommand{\heading}[1]{\textbf{#1}}

\newcommand{\Is}{I^\star}
\newcommand{\If}{I^\star_f}
\newcommand{\Wi}{\mathrm{Wi}}
\newcommand{\WiL}{\mathrm{Wi}_L}
\newcommand{\Isavg}{\langle I^* \rangle}

\begin{document}

\title{
Bioluminescence in turbulence: intermittent straining lights up dinoflagellates
}

\author{Praphul Kumar}
\email{praphul.kr77@gmail.com}
\author{Jason R. Picardo}%
 \email{picardo@iitb.ac.in}
\affiliation{Department of Chemical Engineering, Indian Institute of Technology Bombay, Mumbai 400076, India}%





\begin{abstract}

Dinoflagellates are marine phytoplankton that emit flashes of light in response to flow-induced deformation; they are responsible for illuminating breaking-waves, wakes of ships, and other intensely turbulent spots of the upper ocean. Here, we ask how bioluminescence is affected by the fluctuating nature of turbulence---a question motivated by the dependence of emitted flashes on both the extent and rate of deformation. Introducing a light-emitting dumbbell as a minimal model, we study the Lagrangian dynamics of flashing in a homogeneous isotropic turbulent flow, and contrast it with that in an extensional flow and a Gaussian random flow. We show that turbulent fluctuations strongly enhance bioluminescence, while introducing a Poisson-like stochasticity in the flashing dynamics. Furthermore, the intermittent fluctuations of the velocity-gradient subjects the dinoflagellate to bursts of extreme straining and produces bright flashes---more intense, though less frequent, than what would result from Gaussian fluctuations. Our results suggest that radiant displays of marine bioluminescence are strongly promoted by turbulence and its dissipation-scale intermittency.

\end{abstract}

\maketitle


\heading{Introduction.} 
Dinoflagellates are bioluminescent phytoplankton found in the upper ocean and along shore lines. They emit flashes of light when deformed by external mechanical forces, which could be caused by direct contact with predators or by fluid motion \citep{Latz24-review}. While this deformation-induced emission of light is thought to be a defense mechanism against predation, it also produces spectacular nocturnal displays by lighting up the churning waters below breaking waves and around ships bows \citep{staples1966distribution}. Likened to the northern lights in their mesmerizing beauty \citep{bbc23}, these prominent displays of bioluminescence are clearly associated with turbulent flows. In fact, turbulence is generated artificially by bathyphotometers (devices which measure the in situ bioluminescence of the ocean), in order to elicit an intense luminous response~\citep{WIDDER1993607}. These turbulent motions are certainly associated with large mean-stresses that would strongly deform the dinoflagellates; but does the fluctuating nature of turbulence play a role in the stimulation of luminescence?

Ultimately, light is produced by pH-dependent chemical reactions that occur inside specialized organelles called scintillons; the key reaction is the oxidation of luciferase, catalyzed by the enzyme luciferin~\citep{Valiadi-dino-review}. The deformation of the dinoflagellate triggers an action potential which opens voltage-gated proton channels in the membrane of the scintillons, thereby reducing the pH within them and causing the oxidization reaction~\citep{Valiadi-dino-review}. The process of mechanotransduction (the conversion of the deformation stimulus into an intracellular signal that fires-off the action potential) involves stretch-activated ion channels in the plasma membrane \citep{Latz-stretch}, as well as an increase in the cytoplasmic concentration of $\mathrm{Ca}^2+$ ions that are released from intracellular stores \citep{Latz-Ca}.

The flow-induced deformation of a dinoflagellate, and the associated light emission, depends on the viscous stresses it experiences (experiments have shown that pressure and acceleration do not stimulate bioluminescence \citep{Blaser2002-oscillating,Latz04-converging}). The magnitude of shear stress was shown to be a key factor in producing bioluminescence by Latz and coworkers, in a series of experiments involving dinoflagellates in controlled laminar flows---specifically, cylindrical Couette flow \citep{Latz94-Couette,Maldonado07-shear}, pipe flow \citep{Latz1999-pipe,Rohr2002-threshold-eps}, and converging nozzle flow \citep{Latz04-converging}. The positive correlation between shear stress and illumination is robust, to the extent that dinoflagellate bioluminescence can be used as a means of visualizing the stress field \citep{Latz95-visual2,Rohr2002-threshold-eps,Stokes04-waves}. In transient laminar flows, the rate of change of shear stress was also seen to influence the intensity of illumination, with rapidly increasing stress leading to a bright response. This was directly observed in temporally-developing cylindrical-Couette flow \citep{Latz2005-transient}, while experiments in an oscillating planar-Couette flow \citep{Blaser2002-oscillating} were qualitatively reproduced by a luminescence model \citep{Stokes2005-model} that included a dependence on the rate of change of shear stress. 

In terms of deformation, the experimentally-observed dependence on the magnitude and rate-of-change of stress implies that the brightness of a dinoflagellate's response increases with both the \textit{extent} and \textit{rate} of deformation. This is exactly what has been observed and quantified by \citet{jalaal2020}, who measured the light emitted by the dinoflagellate \textit{Pyrocystis lunula} when subjected to controlled and varying extents and rates of deformation, via indentation by a micropipette. They synthesized their observations into a phenomenological model that predicts the evolution of emitted light intensity given an input rate of deformation. We use this model in the present work.

The dependence of bioluminescence on the rate of change of deformation is of particular relevance in turbulence. Thinking of the dinoflagellate as a sub-Kolmogorov deformable particle, we expect it to experience a fluctuating strain-rate that depends on the velocity-gradient it samples as it is transported in the flow. Over long times, the fluctuating velocity-gradient will act to stretch the dinoflagellate at a time-averaged rate that is proportional to the Lagrangian Lyapunov exponent of the turbulent flow; however, the instantaneous strain-rate will exhibit large deviations from the mean. (The connection between line element stretching and deformation of soft particles has been formalized in the case of an elastic dumbbell by \citet*{bfl00,bfl01} and \citet{c00}). Indeed, the velocity gradient in turbulence exhibits strongly non-Gaussian, intermittent statistics \citep{Ishihara2007,Buaria2019,Buaria2022}, which will result in the dinoflagellate experiencing periods of mild straining interrupted by extreme events of intense straining. 

The possible influence of the intermittent fluctuations of the turbulent velocity gradient on the life of plankton has been recognized, with regard to 
 encounter (collision) rates and nutrient update \citep{Franks22-eps,Franks24-gradv}. Here, we aim to understand how the time-varying deformations produced by the turbulent velocity-gradient impacts the production of light by dinoflagellates. 
 To this end, we combine the deformation-induced bioluminescence model of \citet{jalaal2020} with the simplest mechanical model of a deformable particle, namely the elastic dumbbell. The flashing dynamics of this light-emitting dumbbell in a homogeneous isotropic turbulent flow is examined, for a range of turbulent dissipation rates. A comparison with a uniform extensional flow, which is devoid of fluctuations, and a Gaussian flow, which is stochastic but \textit{not} intermittent, suggests that intermittent fluctuating straining of dinoflagellates plays an important role in radiant displays of marine bioluminescence.

\begin{figure*}
\centering
\includegraphics[width=0.32\textwidth]{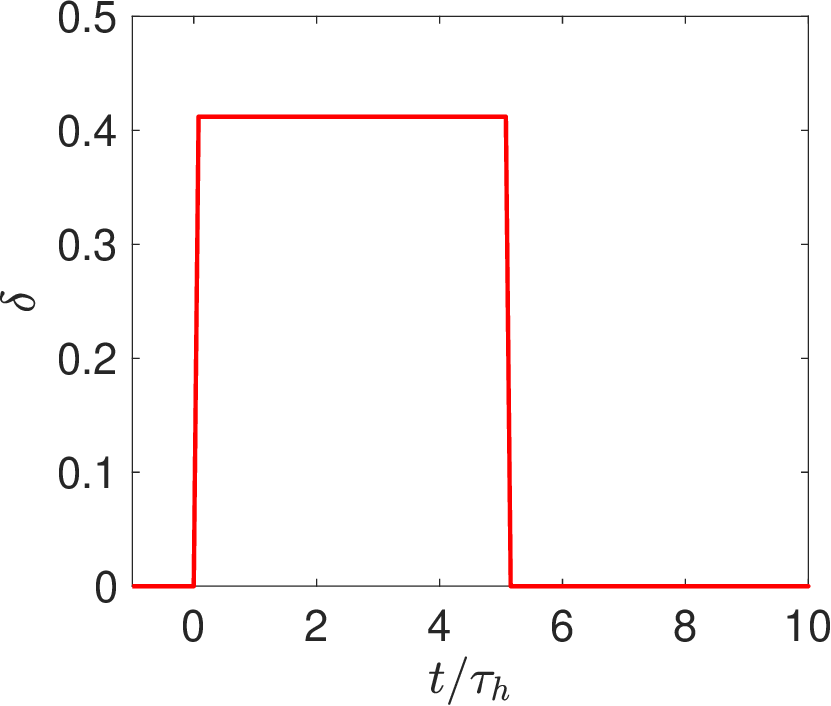}
\put (-19,30){(a)}
\hspace{0.01\textwidth}
\includegraphics[width=0.32\textwidth]{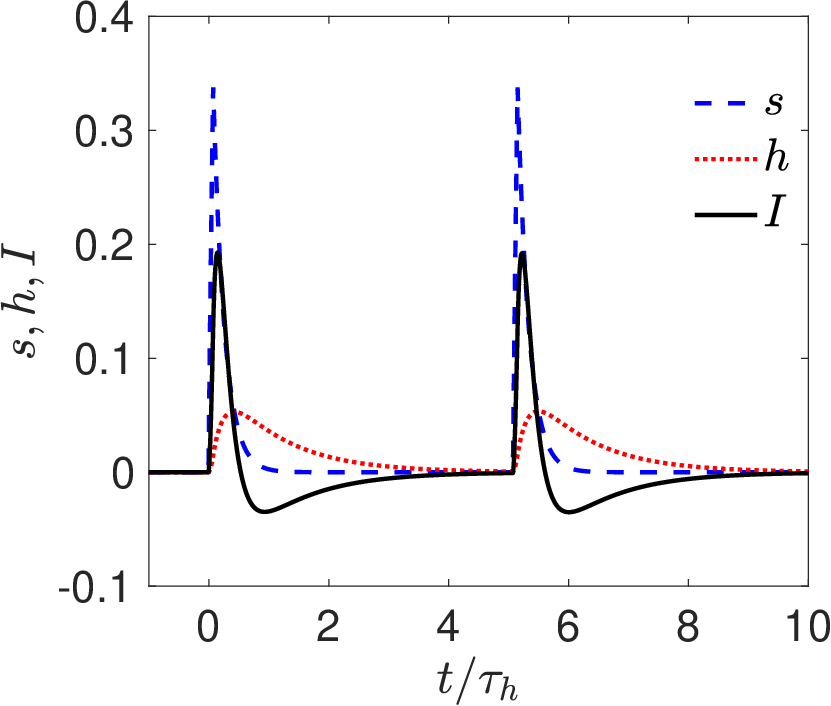}
\put (-19,30){(b)}
\hspace{0.01\textwidth}
\includegraphics[width=0.32\textwidth]{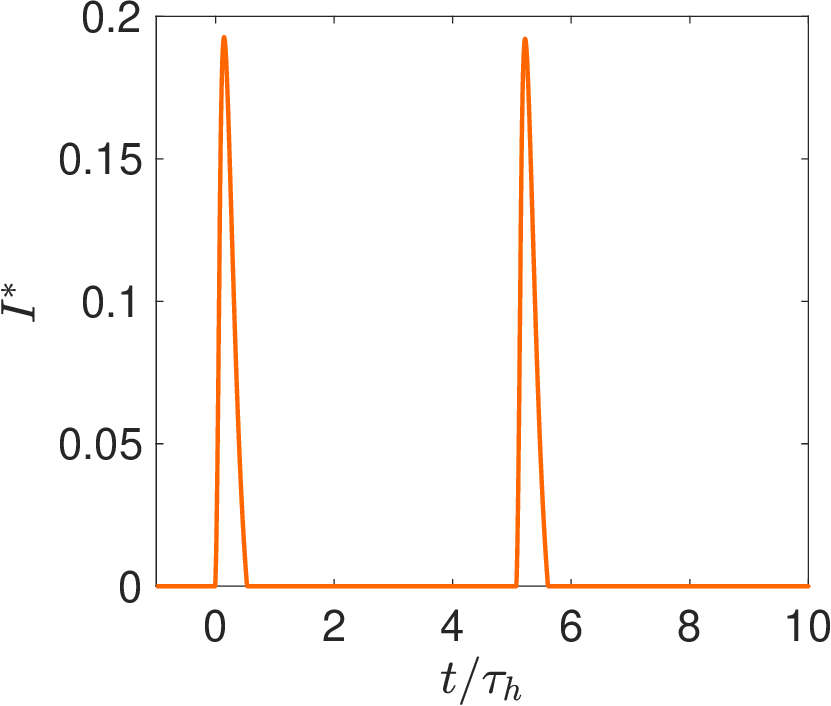}
\put (-19,30){(c)}

\caption{\label{fig:light-model} Illustration of the light-emission model in Eq.~\eqref{eq:light}. (a) Strain profile ${\delta}(t)$, with rate of change of strain $\dot \delta = \tau_s^{-1} \approx 5 \,\tau_h^{-1}$ during deformation and relaxation. (b) Evolution of $s$, $h$, and $I$. (c) Intensity of light $I^*$ obtained by replacing negative values of $I$ with zero (no light emission).}
\end{figure*}

\heading{Light emission in response to deformation.}
In the phenomenological model of \citet{jalaal2020}, the applied strain $\delta(t)$ (defined below, in terms of the elastic dumbbell) generates an intra-cellular signal $s(t)$ (such as a flux of $\mathrm{Ca}^{2+}$ ions), which in turn triggers the emission of light with an intensity that is represented by $I(t)$. The cell's internal relaxation processes by which the flash of light ends, and the cell resets for a subsequent flash, are represented by $h(t)$. The dynamics of these three variables are associated with time-scales $\tau_s$, $\tau_I$, and $\tau_h$:
\begin{equation}
    \dot{s} + \tau_{s}^{-1} s = |\dot{\delta}|,\quad
    \tau_I \dot{I} = s - h - I, \quad
    \tau_h \dot{h} = s - h.\label{eq:light}
\end{equation}
The equation for $s(t)$, inspired by Maxwell's viscoelastic model, accounts for the dependence of light emission on both the extent and rate of deformation \citep{jalaal2020}.
The three time-scales were determined by fitting the model to data from experiments, performed for different deformation rates $\dot\delta(t)$, maintained for different time intervals (thus yielding different extents of deformation): $\tau_{s} = 0.027$~s, $\tau_I = 0.012$~s, $\tau_h = 0.14$~s. 
\citet{jalaal2020} observed flashes not only when the dinoflagellate was deformed by the pipette but also when the cell relaxed as the pipette was withdrawn. Therefore, we use the absolute value of $\dot\delta(t)$ in the model. 
Figure~\ref{fig:light-model} illustrates the behavior of the model for one deformation-relaxation cycle. The imposed deformation is shown in Fig.~\ref{fig:light-model}(a) and the response of all three variables are presented in Fig.~\ref{fig:light-model}(b). An artifact of the simplicity of the model is that, depending on the form of the input signal $\dot\delta(t)$, $I$ can attain negative values after a flash is completed; we observe this in Fig.~\ref{fig:light-model}(b). Since such negative values have no physical meaning, we focus on just the positive values and consider $\Is = I$ for $I \ge 0$ and $\Is = 0$ for $I < 0$. The final output light signal is presented in Fig.~\ref{fig:light-model}(c).

As noted by \citet{jalaal2020}, this model does not account for deformation thresholds, neither for the onset nor for the saturation of light emission, both of which have been observed in experiments \citep{Latz04-4species-threshold,Latz1999-pipe,jalaal2020}. We therefore restrict our study to a range of turbulent dissipation rates for which the fluid stress within a Kolmogorov eddy lies within experimentally-observed stress thresholds (discussed further below). 

When a dinoflagellate flashes repeatedly in quick succession, the intensity of light decreases with each flash \citep{jalaal2020,Stokes2005-model} (The model in Eq.~\eqref{eq:light} was developed by \citet{jalaal2020} for just the first flash.) The illumination of a local zone of intense turbulence in the ocean (beneath breaking waves or near a ships bow) would be sustained despite this decay because of the entrainment of fresh dinoflagellates into the region of strong fluctuating stresses. For simplicity, we ignore both decay and entrainment in this work, with the expectation that these counteracting factors will not alter our conclusions regarding the qualitative effect of turbulent straining.


\heading{Deformation in a turbulent flow.}
Dinoflagellates have cell walls that resist deformation, though the cytoplasm also plays a role in maintaining the cell's structure \citep{jalaal2020}. Here we use the simplest mechanical model for an elongated deformable object---the elastic dumbbell. The end-to-end extension vector $\bm R$ is governed by
\begin{equation}
\dot{\bm {R}} = \nabla {\bm v}({\bm x}_{cm}(t),t) \cdot \bm{R} - f(\bm{R}) \left( \frac{R - R_{eq}}{R} \right)\frac{\bm{R}}{2\tau_E},
\quad \mathrm{with} \quad
f(\bm{R}) = \left[1 - \left({R}/{R_{m}}\right)^2\right]^{-1},
\label{eq:dumbbell}
\end{equation}
where $R = |\bm R|$ is the instantaneous extension, while \(R_{eq}\) is the equilibrium extension, to which the dumbbell relaxes in a quiescent fluid. The time-scale of relaxation is \(\tau_E\). The dumbbell has a maximum extension \(R_m\), which is enforced by the finitely extensible nonlinear elastic (FENE) force coefficient \(f(R)\). 

The velocity gradient of the turbulent flow \(\nabla{\bm v}\) is evaluated at the position of the center of mass of the dumbbell, ${\bm x}_{cm}$, which evolves like a tracer in the flow satisfying $\dot{{\bm x}}_{cm} = \bm v({\bm x}_{cm})$. We ignore the weak motility of the dinoflagellate and its slight negative buoyancy \citep{jalaal2020}, properties that would prevent it from behaving exactly like a tracer.

\begin{figure*}
\centering
\includegraphics[width=0.325\textwidth]{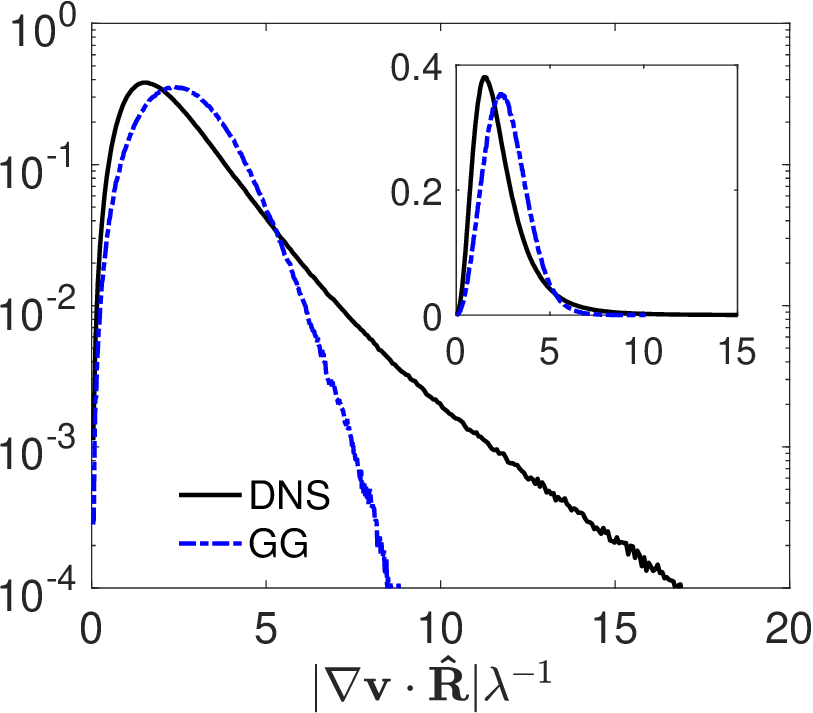}
\put (-19,30){(a)}
\hspace{0.01\textwidth}
\includegraphics[width=0.315\textwidth]{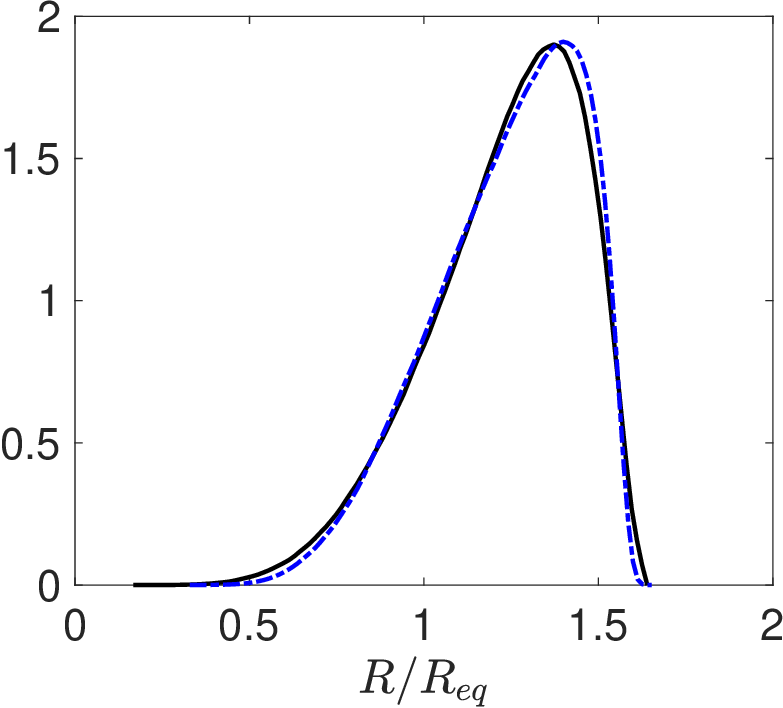}
\put (-19,30){(b)}
\put (-110,111) {$\Wi = 0.3$}
\hspace{0.01\textwidth}
\includegraphics[width=0.325\textwidth]{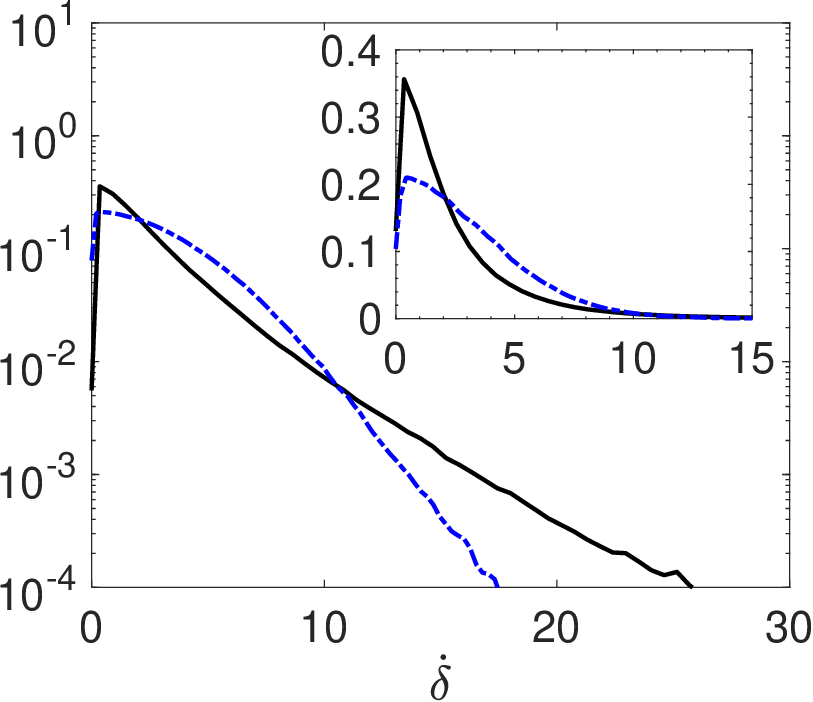}
\put (-19,30){(c)}
\put (-130,30) {$\Wi = 0.3$}

\caption{\label{fig:deformation} Comparison of the PDFs of (a) strain-rates experienced by the dumbbell $|\nabla \bm{v} \cdot \hat{\bm{R}}|\lambda^{-1}$ (where $\hat{\bm R} = \bm R/R$), (b) dumbbell extension $R$, and (c) rate of change of strain $\dot \delta = \dot {R}/R_{eq}$, in the turbulent DNS and in the Gaussian gradient (GG) model. Here $\Wi = 0.3$; analogous results are obtained for $\Wi = 1.0$ \citep{supplement}.}
\end{figure*}

In our simulations, we use a dataset of \(\nabla{\bm v}\), calculated along $10^4$ tracer trajectories in a direct numerical simulation (DNS) of a homogeneous isotropic turbulent flow, with a Taylor Reynolds number $Re_\lambda = 111$ \citep{jr17}. The DNS is performed in a tri-periodic cube with a $512^3$ pseudo-spectral grid; see the supplementary material (SM) for further details \citep{supplement}. This Lagrangian dataset has been used previously for studying the stretching of polymers (dumbbells and chains) in turbulence \citep{ppv23,gvpp25}. The stretching action of the fluctuating velocity gradient is best characterized by the Lyapunov exponent along tracer trajectories $\lambda$, which is calculated from the data for \(\nabla{\bm v}\) using the continuous $QR$ method \citep{drv97}. $\lambda$ is the long-time exponential stretching rate of line elements in the flow. We use $\lambda$ to define the Weissenberg number of the dumbbell as $\Wi = \lambda \tau_E$. 

To understand the effect of turbulent fluctuations, we perform simulations in a uniform extensional flow,
\(
\nabla {\bm v} = \bigl(\begin{smallmatrix} \dot \gamma & 0 \\ 0 & -\dot \gamma \end{smallmatrix}\bigr)
\), 
with strain-rate $\dot \gamma = \lambda$. To isolate the effect of extreme-valued fluctuations, we also consider a \textit{Gaussian} time-correlated random model for \(\nabla{\bm v}\), specifically that of \citet{bkl98}. We set the parameters of this model such that it yields the same Lyapunov exponent as the turbulent flow, as well as approximately the same correlation time-scales of vorticity and strain-rate (see the SM for details \citep{supplement}). Figure~\ref{fig:deformation}(a) compares the probability distribution function (PDF) of the magnitude of the strain-rate experienced by the dumbbell, $|\nabla \bm{v} \cdot \bm{\hat{R}}|$ (where $\hat{\bm R} = \bm R/R$), in the turbulent flow (DNS) to that in the random flow with Gaussian gradients (GG). Clearly, the intermittent velocity gradient in the turbulent flow subjects the dumbbell to more episodes of very strong and very weak strain-rates, but less episodes of moderate strain-rates. 

Having a Lagrangian time-series for \(\nabla{\bm v}\), we integrate \eqref{eq:dumbbell} in time using a fourth-order Runge-Kutta method. From $R(t)$, we obtain the strain as $\delta(t) = ({R(t) - R_{eq}})/{R_{eq}}$; its time-derivative $\dot \delta$ is the input to the luminescence model \eqref{eq:light}.
We set
\(R_{eq} = 1\), \(R_m = 1.7R_{eq}\), and $\Wi = 0.3$. These choices yield a mean strain of $\sim 0.4$ which is comparable to that in the experiments of \cite{jalaal2020}. We have checked that our conclusions are not sensitive to the precise choice of these parameters, as demonstrated by additional results for $\Wi = 1.0$ in the SM \citep{supplement}. 

The PDF of the extension is shown in Fig.~\ref{fig:deformation}(b) for the DNS and the GG flow. The results are almost the same; as shown by \citet{ppv23}, the statistics of dumbbell extension are largely determined by the time-averaged rate of line-element stretching, that is by $\lambda$, which is the same in the two flows. So, similar extension can result from sustained moderate-straining (GG) as from relatively weak straining interspersed by bursts of strong straining (DNS) \citep{ppv23}. In contrast, the PDF of the rate of change of extension $\dot R/R_{eq}$, which equals $\dot \delta$, is sensitive to the presence of extreme velocity-gradient fluctuations in the turbulent flow, as demonstrated by Fig.~\ref{fig:deformation}(c). Hence, the dependence of luminescence on the rate of deformation $\dot \delta$ must render the flashing dynamics sensitive to extreme-valued turbulent fluctuations. 

We analyze the flashing dynamics for a range of \textit{luminous} Weissenberg numbers, defined as $\WiL = \lambda\,\tau_h$ (recall that $\tau_h$ is the time-scale over which the cell resets itself after flashing). This is equivalent to considering a range of turbulent dissipation rates $\epsilon$, since $\lambda \sim \epsilon^{1/2}$ as discussed below. Now, dinoflagellates emit light only once the stress experienced by them exceeds a lower threshold \citep{Latz24-review}, which is species-specific and was shown by \citet{Latz04-4species-threshold}, in a study of four different species, to vary from 0.02 to 0.3 \si{N.m^{-2}}. Experiments have also found an upper threshold of $\approx 1$ \si{N.m^{-2}} beyond which light emission saturates \citep{Latz1999-pipe}. The stress at the Kolmogorov scale of the flow may be estimated as $\mu \tau_\eta^{-1}$ where $\tau_\eta = (\mu/\rho\epsilon)^{1/2}$ is the Kolmogorov time-scale, which is an estimate of both the strain-rate within the smallest eddy and of its turnover time \citep{f95} (here, $\mu$ and $\rho$ are the liquid viscosity and density). In water, a Kolmogorov-scale stress in the range 0.02 to 1 \si{N.m^{-2}} requires a turbulent dissipation rate $\epsilon$ between $4 \times 10^{-4}$ and $1$ \si{m^2.s^{-3}}. The corresponding $\tau_\eta$ lies between $5 \times 10^{-2}$ and $10^{-3}$ \si{s}. Since in fully developed turbulence, one typically has $\lambda \sim 0.1 \,\tau_\eta^{-1}$ \citep{bbbcmt06} (for our DNS data the exact relation is $\lambda \sim 0.136 \,\tau_\eta^{-1}$ \citep{ppv23}), we estimate that $\WiL \sim 0.1$ at the onset of luminescence while increasing $\WiL$ beyond $\sim 10$ will not increase light production. We therefore restrict our study to $0.1 < \WiL < 10$. 

The Kolmogorov length scale $\eta = \epsilon^{-1/4} (\mu/\rho)^{3/4}$ for the dissipation rates of interest spans $\sim 200$ to $\sim30$ \si{\mu.m}. And most dinoflagellates have sizes $\sim 100$ \si{\mu.m} \citep{jalaal2020,Latz04-4species-threshold}. So, for the strongest turbulent flows, the dinoflagellate can extend beyond the Kolmogorov scale \citep{Latz1999-pipe}, though not by much. In such cases, the velocity difference across the dinoflagellate is not determined entirely by $\nabla \bm v$, but will have corrections from higher-order terms in the Taylor series. Accounting for super-Kolmogorov sizes is one of several directions for future work, which are discussed towards the end of this paper.

Note that background turbulence in the upper ocean has much smaller values of turbulent dissipation ($10^{-10}$ to $10^{-5}$ \si{m^2.s^{-3}} \citep{Franks22-eps}) than that required for dinoflagellate bioluminescence. Hence, bright displays of light occur only in zones of abnormally intense turbulence, such as breaking waves and ship wakes \citep{Rohr2002-threshold-eps,Stokes04-waves}.

\begin{figure*}
\centering
\includegraphics[width=0.32\textwidth]{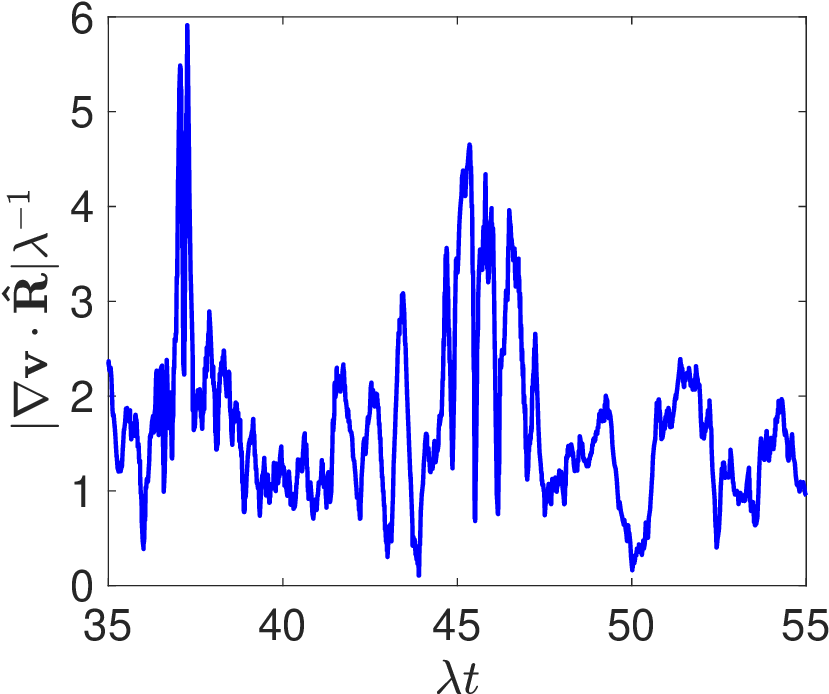}
\put (-19,111){(a)}
\hspace{0.01\textwidth}
\includegraphics[width=0.32\textwidth]{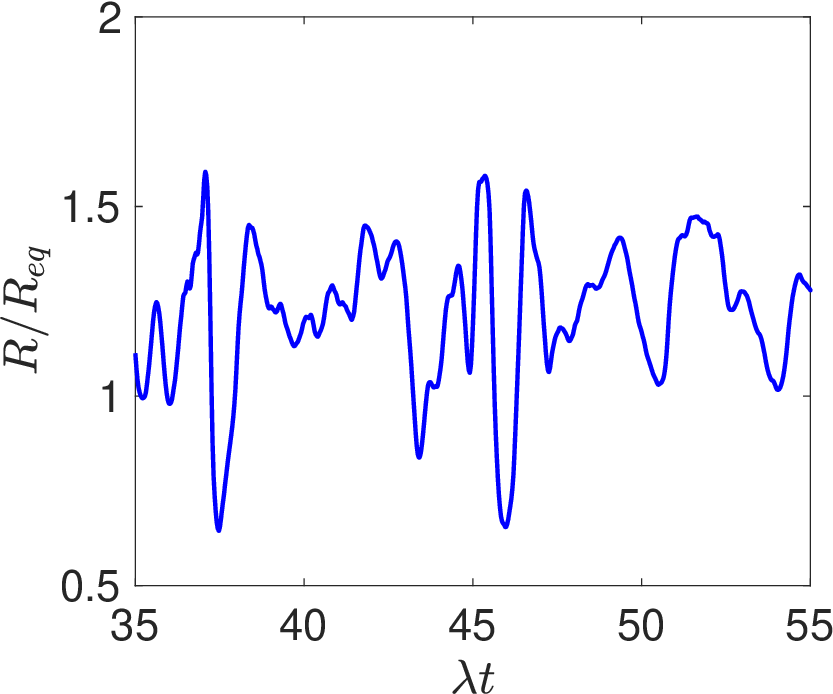}
\put (-19,111){(b)}
\hspace{0.01\textwidth}
\includegraphics[width=0.32\textwidth]{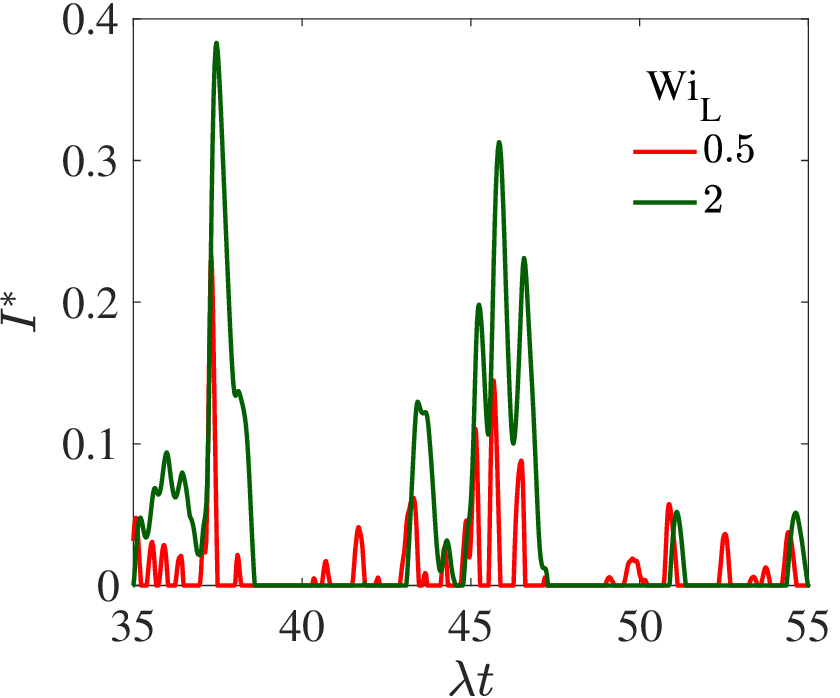}
\put (-19,111){(c)}
\caption{\label{fig:trace-turb} 
Evolution of (a) the magnitude of the strain-rate experienced by the dumbbell $|\nabla \bm{v} \cdot \hat{\bm R}|$ (where $\hat{\bm R} = {\bm R}/R$), (b) the end-to-end extension $R$, and (c) the light intensity $I^*$ (for $\WiL = 0.5$ and $2.0$), in the turbulent flow. These repeated flashes may be contrasted with the single, much weaker, flash in an extension flow (see Fig~S1 in the SM \citep{supplement}). Here $\Wi = 0.3$.}
\end{figure*}

\heading{Flashing in turbulence.} The light-emitting dumbbell (Eqs.~\eqref{eq:light} and \eqref{eq:dumbbell}) quickly attains a statistically stationary state in the turbulent flow (after a few multiples of $\tau_h$) and exhibits a fluctuating light response $\Is$ that consists of apparently random flashes. Typical time-traces of the strain-rate along the dumbbell $|\nabla \bm{v} \cdot \hat{\bm R}|$, the extension $R$, and the light intensity $\Is$ are illustrated in Figs.~\ref{fig:trace-turb}(a-c). 
The strongest flashes in Fig.~\ref{fig:trace-turb}(c) occur during the strongest straining events in Fig.~\ref{fig:trace-turb}(a) (around $\lambda t$ = 37 and 46); this is to be expected because the strain-rate $\nabla \bm{v} \cdot {\bm R}$ directly affects the rate of deformation $\dot \delta = \dot R/R_{eq}$ [see Eq.~\eqref{eq:dumbbell}], which in turn drives the internal cellular signal $s$ [see Eq.~\eqref{eq:light}]. The extension indeed undergoes its largest fluctuations near the times of the strongest flashes [Fig.~\ref{fig:trace-turb}(b)].

\begin{figure*}
\centering
\includegraphics[width=0.33\textwidth]{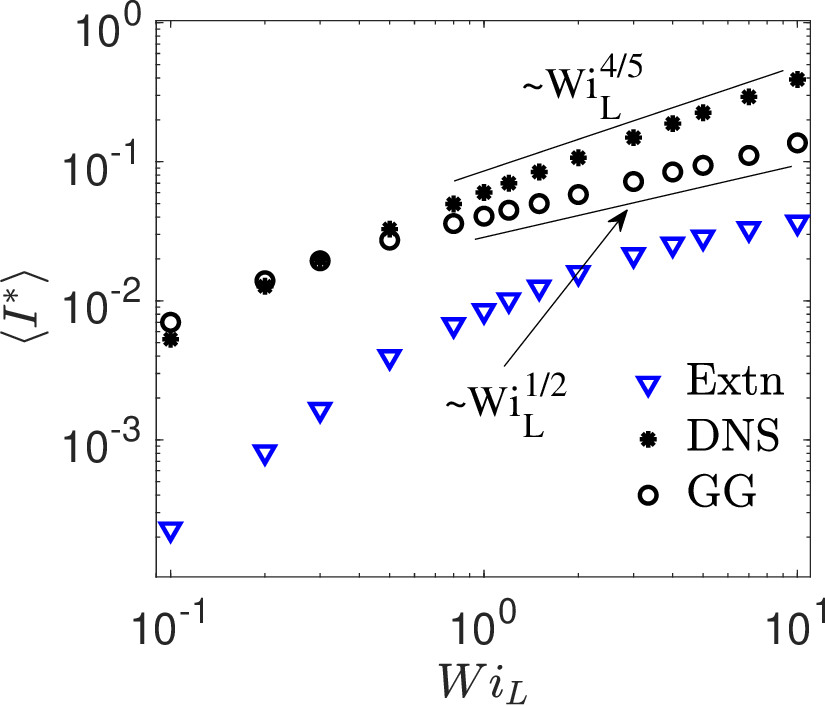}
\put (-118,111){(a)}
\includegraphics[width=.325\textwidth]{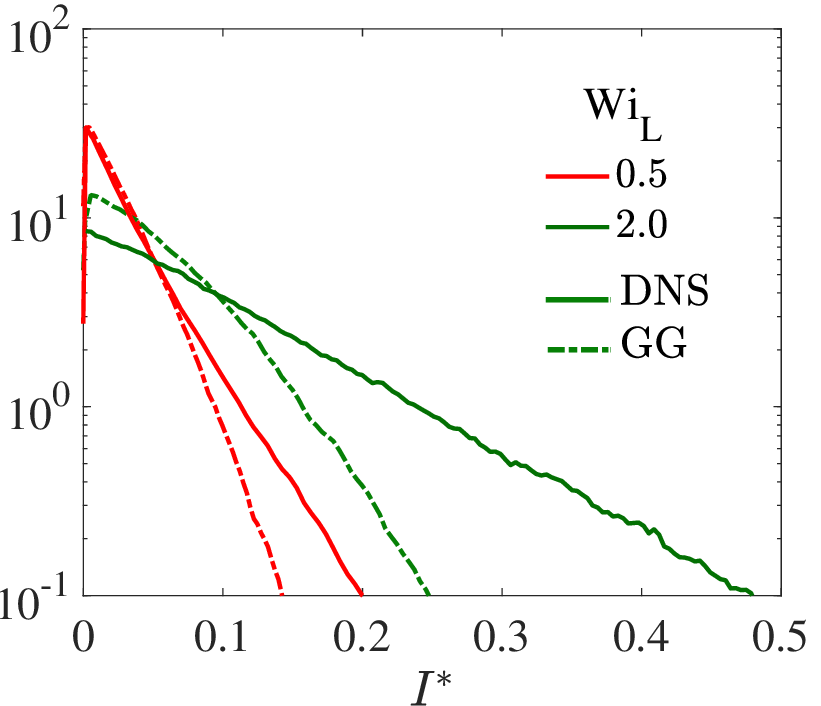}
\put (-118,111){(b)}
\includegraphics[width=0.32\textwidth]{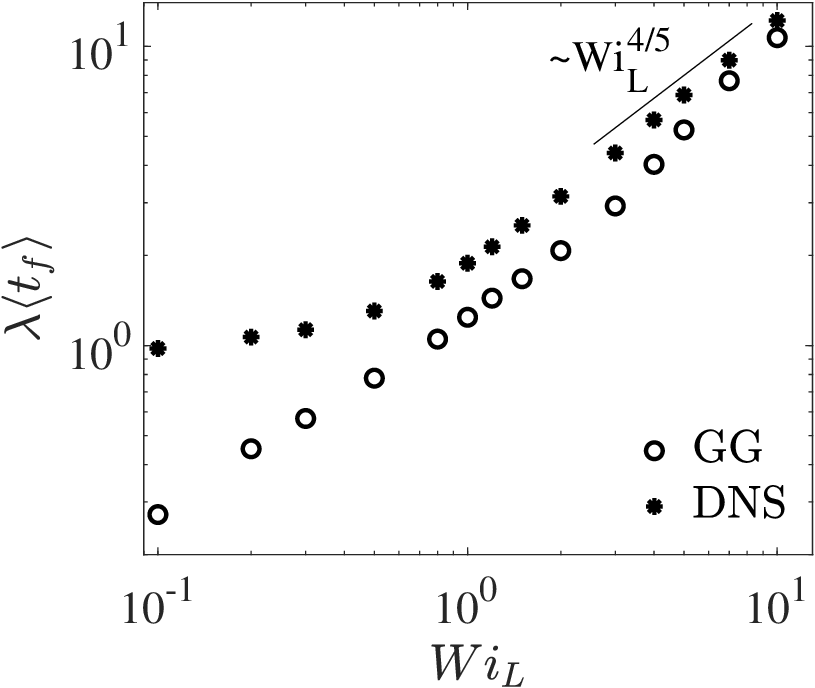}
\put (-118,111){(c)}

\caption{\label{fig:I} (a) Variation of the mean light intensity $\langle I^* \rangle$ with $\WiL$, in the turbulent DNS and the Gaussian-gradient (GG) flow. (b)  PDF of the light intensity $\Is$, for $\WiL = 0.5$ and $2.0$, in the two fluctuating flows. (c) Variation of the mean waiting-time between flashes $\langle t_f\rangle$ with $\WiL$. Here, $\Wi = 0.3$; analogous results are obtained for $\Wi = 1.0$ \citep{supplement}.}
\end{figure*}

The results for different $\WiL$ in Fig.~\ref{fig:trace-turb}(c) show that larger values of $\WiL$, which correspond to stronger turbulent flows with greater dissipation rates, result in brighter flashes. This is confirmed by Fig.~\ref{fig:I}(a), which presents the mean light intensity $\Isavg$ (averaged over all trajectories and over time in the stationary state) for various $\WiL$. Results are presented for the turbulent DNS, the random Gaussian flow, and an extensional flow. Focusing first on the DNS results, we observe an approximate linear increase of the mean light intensity with $\WiL$. This is consistent with the experiments in turbulent pipe flow of \citet{Rohr2002-threshold-eps} and \citet{Blaser2002-oscillating}, who both report an approximately linear increase of light intensity with the magnitude of the wall shear stress (the dinoflagellates light up near the wall and so the wall shear stress was used as an estimate of the stress experienced by them). Recall that $\WiL$ is proportional to the viscous stress experienced by the dumbbell, since $\WiL \propto \lambda \propto \tau_\eta^{-1}$  and $\mu \tau_\eta^{-1}$ is the Kolmogorov-scale viscous stress.

\heading{Intermittent fluctuations brighten flashes.}
To understand the role of flow fluctuations, we calculated the light intensity emitted due to deformation in a uniform extensional flow, with a strain-rate that matches the Lyapunov exponent of the turbulent flow. The dumbbell's extension initially increases and then reaches a constant value, as the dumbbell aligns with the stretching direction of the extensional flow. As a consequence there is only one flash (see Fig.~S1 in the SM \citep{supplement}), the intensity of which varies depending on the initial orientation of the dumbbell (qualitatively similar results are obtained in planar shear flow, wherein the dumbbell aligns with the flow direction). On averaging $\Is$ over the initial orientation, and over the time duration of the flash (i.e., when $\Is \neq 0$), we obtain the results for $\Isavg$ shown in Fig.~\ref{fig:I}(a) (inverted triangles). Though the average straining rate is comparable in the two flows (since $\dot \gamma = \lambda$), the fluctuating turbulent flow elicits a much brighter luminous response than the extensional flow.

The effect of intermittent velocity-gradient fluctuations are revealed by comparing the results in the turbulent flow (DNS) with those in the random flow with Gaussian gradients (GG). The mean light response is seen to be greater in the DNS at large $\Wi_L$, though the difference appears to become negligible at small $\WiL$ [see Fig.~\ref{fig:I}(a)]. However, on comparing the PDFs of $\Is$ in the two flows [Fig.~\ref{fig:I}(b)], we find that the flashing dynamics is sensitive to extreme velocity-gradient fluctuations even at small $\WiL$: the PDF of $\Is$ shows much wider tails in the turbulent flow. Given that $\Isavg$ is the same at small $\WiL$, the wider tails imply that the flashes in the intermittent turbulent flow are brighter but less frequent than those in the Gaussian random flow.

\heading{Waiting time between flashes.} To quantify how the flashing frequency depends on the strength of the turbulent flow, we calculate the waiting time between flashes $t_f$. We identify flashes as the periods of time during which $\Is \ge \If$, where the threshold $\If = \alpha \Is_{max}$ (the maximum value varies with $\WiL$ and is calculated over the whole ensemble of trajectories and across time in the stationary state). We set $\alpha = 0.15$ after checking that our conclusions are not sensitive to the precise value of the threshold. The waiting time $t_f$ is then given by the intervals of time for which $\Is < \If$ [see Fig.~S2(a) in the SM \citep{supplement}]. The PDF of $t_f$ is found to have approximately exponential tails [Fig.~S2(b) in \citep{supplement}], 
showing that the fluctuations of the flow introduce an apparent stochasticity into the light dynamics, such that the flashes may be modeled as a Poisson process.
The mean waiting time $\langle t_f \rangle$ is presented as a function of $\WiL$ in Fig.~\ref{fig:I}(c) for the DNS and the GG flow. As anticipated, the flashes are less frequent ($\langle t_f \rangle$ is larger) in the intermittent turbulent flow, especially at small $\WiL$. As the turbulence becomes more intense the rate of flashing increases and becomes insensitive to the presence of extreme velocity-gradient fluctuations. The extreme gradients always generate larger-amplitude flashes, however, as evidenced by the difference in $\Isavg$ growing with $\WiL$ [Fig.~\ref{fig:I}(a)] even as the difference in $\langle t_f \rangle$ shrinks [Fig.~\ref{fig:I}(c)].

\heading{Concluding remarks.}
With a minimal model of a flashing dinoflagellate that accounts for the dependence of light emission on the extent and rate of deformation, we have shown how the fluctuating and intermittent nature of small-scale turbulent straining enhances bioluminescence. Experiments in turbulent pipe flow by \citet{Latz1999-pipe} showed that the mean light intensity emitted by dinoflagellates near the wall is the same in a laminar and turbulent flow provided they have the same mean wall shear-stress, which is equivalent to having the same mean wall shear-rate or strain-rate. However, in turbulence, the Eulerian mean at a fixed location is not the same as the Lagrangian mean along the path of a dinoflagellate. 
The Lagrangian mean will be reduced by cross-pipe turbulent mixing, which will prevent dinoflagellates from persistently experiencing the high strain-rates near the wall. The fact that the light intensity remains the same as the corresponding laminar flow could be attributed to the mechanism uncovered here---the lower Lagrangian mean strain-rate is compensated for by turbulent fluctuations which enhance light emission. Simulations of our light-emitting dumbbell in a turbulent pipe flow should help clarify how bioluminescence varies across the laminar-turbulent transition. 

We have focused here on Lagrangian dynamics and ignored the spatial distribution of luminescence. When exploring the pattern of illumination, it would be relevant to extend the mechanical model to account for the 
motility and possible super-Kolmogorov size of dinoflagellates. These properties cause deviations from tracer-like transport and result in preferential sampling of the flow---
microswimmers sample non-vortical regions~\citep{Pujara2018}, while long rods sample vortical regions~\citep{Picardo2020}. This would not only alter the statistics of the strain-rates experienced by dinoflagellates but also lead to clustering~\citep{Wilczek2018}. One could also examine how the shape of a dinoflagellate (rod-like or spherical) affects bioluminescence, by replacing the elastic dumbbell with a more complex structural model like a soft ellipsoid~\citep{Gao2011,Nott2022}.

The process of light emission may also be modeled in more detail, as done by \citet{Stokes2005-model} (see also \citep{Stokes04-waves,Hao24}) who use an empirical form for the flash and model flash-triggering as a Poisson process with a firing rate that depends on the fluid stress experienced by the dinoflagellate (the magnitude and the rate of change). This model accounts for the stress-thresholds for light response and the stochasticity of internal cell processes (though the latter is not crucial in turbulence, since even without internal randomness the external turbulent flow produces Poisson-like stochastic flashing). Reformulating this model in terms of deformation and applying it to a turbulent flow, while accounting for cell fatigue (decay of repeated flashes) and turbulent entrainment of fresh cells, would be an important step towards quantitative prediction of turbulent bioluminescence. \\

\heading{Acknowledgments.} We thank Samriddhi Sankar Ray (ICTS, Bengaluru) for sharing the Lagrangian trajectory database. This work was supported by the Indo–French Centre for the Promotion of Advanced Scientific Research (IFCPAR/CEFIPRA, project no. 6704-1). P.K. is grateful for funding from IIT Bombay under the Institute Postdoctoral Fellowship program. J.R.P. acknowledges his Associateship with the International Centre for Theoretical Sciences (ICTS), Tata Institute of Fundamental Research, Bengaluru,
India. We are thankful for the computational resources provided by the National PARAM Supercomputing Facility PARAM SIDDHI-AI at CDAC, Pune, India. 
Simulations were also performed on the IIT Bombay workstation \textit{Gandalf} (procured through DST-SERB grant SRG/2021/001185).

\FloatBarrier
\bibliography{biolumin}

\providecommand{\noopsort}[1]{}\providecommand{\singleletter}[1]{#1}%
\begin{thebibliography}{41}%
\makeatletter
\providecommand \@ifxundefined [1]{%
 \@ifx{#1\undefined}
}%
\providecommand \@ifnum [1]{%
 \ifnum #1\expandafter \@firstoftwo
 \else \expandafter \@secondoftwo
 \fi
}%
\providecommand \@ifx [1]{%
 \ifx #1\expandafter \@firstoftwo
 \else \expandafter \@secondoftwo
 \fi
}%
\providecommand \natexlab [1]{#1}%
\providecommand \enquote  [1]{``#1''}%
\providecommand \bibnamefont  [1]{#1}%
\providecommand \bibfnamefont [1]{#1}%
\providecommand \citenamefont [1]{#1}%
\providecommand \href@noop [0]{\@secondoftwo}%
\providecommand \href [0]{\begingroup \@sanitize@url \@href}%
\providecommand \@href[1]{\@@startlink{#1}\@@href}%
\providecommand \@@href[1]{\endgroup#1\@@endlink}%
\providecommand \@sanitize@url [0]{\catcode `\\12\catcode `\$12\catcode `\&12\catcode `\#12\catcode `\^12\catcode `\_12\catcode `\%12\relax}%
\providecommand \@@startlink[1]{}%
\providecommand \@@endlink[0]{}%
\providecommand \url  [0]{\begingroup\@sanitize@url \@url }%
\providecommand \@url [1]{\endgroup\@href {#1}{\urlprefix }}%
\providecommand \urlprefix  [0]{URL }%
\providecommand \Eprint [0]{\href }%
\providecommand \doibase [0]{https://doi.org/}%
\providecommand \selectlanguage [0]{\@gobble}%
\providecommand \bibinfo  [0]{\@secondoftwo}%
\providecommand \bibfield  [0]{\@secondoftwo}%
\providecommand \translation [1]{[#1]}%
\providecommand \BibitemOpen [0]{}%
\providecommand \bibitemStop [0]{}%
\providecommand \bibitemNoStop [0]{.\EOS\space}%
\providecommand \EOS [0]{\spacefactor3000\relax}%
\providecommand \BibitemShut  [1]{\csname bibitem#1\endcsname}%
\let\auto@bib@innerbib\@empty
\bibitem [{\citenamefont {Letendre}\ \emph {et~al.}(2024)\citenamefont {Letendre}, \citenamefont {Twardowski}, \citenamefont {Blackburn}, \citenamefont {Poulin},\ and\ \citenamefont {Latz}}]{Latz24-review}%
  \BibitemOpen
  \bibfield  {author} {\bibinfo {author} {\bibfnamefont {F.}~\bibnamefont {Letendre}}, \bibinfo {author} {\bibfnamefont {M.}~\bibnamefont {Twardowski}}, \bibinfo {author} {\bibfnamefont {A.}~\bibnamefont {Blackburn}}, \bibinfo {author} {\bibfnamefont {C.}~\bibnamefont {Poulin}},\ and\ \bibinfo {author} {\bibfnamefont {M.~I.}\ \bibnamefont {Latz}},\ }\bibfield  {title} {\bibinfo {title} {A review of mechanically stimulated bioluminescence of marine plankton and its applications},\ }\bibfield  {journal} {\bibinfo  {journal} {Frontiers in Marine Science}\ }\textbf {\bibinfo {volume} {10}},\ \href {https://doi.org/10.3389/fmars.2023.1299602} {10.3389/fmars.2023.1299602} (\bibinfo {year} {2024})\BibitemShut {NoStop}%
\bibitem [{\citenamefont {Staples}(1966)}]{staples1966distribution}%
  \BibitemOpen
  \bibfield  {author} {\bibinfo {author} {\bibfnamefont {R.}~\bibnamefont {Staples}},\ }\href@noop {} {\emph {\bibinfo {title} {The Distribution and Characteristics of Surface Bioluminescence in the Oceans}}},\ Technical report\ (\bibinfo  {publisher} {U.S. Naval Oceanographic Office},\ \bibinfo {year} {1966})\BibitemShut {NoStop}%
\bibitem [{\citenamefont {Grice}(2023)}]{bbc23}%
  \BibitemOpen
  \bibfield  {author} {\bibinfo {author} {\bibfnamefont {N.}~\bibnamefont {Grice}},\ }\href@noop {} {\bibinfo {title} {Bioluminescent plankton: 'it's the northern lights of the ocean'}},\ \bibinfo {howpublished} {\url{https://www.bbc.com/news/uk-wales-65984861}} (\bibinfo {year} {2023}),\ \bibinfo {note} {{BBC News}. Accessed 08-Sep-2025.}\BibitemShut {Stop}%
\bibitem [{\citenamefont {Widder}\ \emph {et~al.}(1993)\citenamefont {Widder}, \citenamefont {Case}, \citenamefont {Bernstein}, \citenamefont {MacIntyre}, \citenamefont {Lowenstine}, \citenamefont {Bowlby},\ and\ \citenamefont {Cook}}]{WIDDER1993607}%
  \BibitemOpen
  \bibfield  {author} {\bibinfo {author} {\bibfnamefont {E.}~\bibnamefont {Widder}}, \bibinfo {author} {\bibfnamefont {J.}~\bibnamefont {Case}}, \bibinfo {author} {\bibfnamefont {S.}~\bibnamefont {Bernstein}}, \bibinfo {author} {\bibfnamefont {S.}~\bibnamefont {MacIntyre}}, \bibinfo {author} {\bibfnamefont {M.}~\bibnamefont {Lowenstine}}, \bibinfo {author} {\bibfnamefont {M.}~\bibnamefont {Bowlby}},\ and\ \bibinfo {author} {\bibfnamefont {D.}~\bibnamefont {Cook}},\ }\bibfield  {title} {\bibinfo {title} {A new large volume bioluminescence bathyphotometer with defined turbulence excitation},\ }\href {https://doi.org/https://doi.org/10.1016/0967-0637(93)90148-V} {\bibfield  {journal} {\bibinfo  {journal} {Deep Sea Research Part I: Oceanographic Research Papers}\ }\textbf {\bibinfo {volume} {40}},\ \bibinfo {pages} {607} (\bibinfo {year} {1993})}\BibitemShut {NoStop}%
\bibitem [{\citenamefont {Valiadi}\ and\ \citenamefont {Iglesias-Rodriguez}(2013)}]{Valiadi-dino-review}%
  \BibitemOpen
  \bibfield  {author} {\bibinfo {author} {\bibfnamefont {M.}~\bibnamefont {Valiadi}}\ and\ \bibinfo {author} {\bibfnamefont {D.}~\bibnamefont {Iglesias-Rodriguez}},\ }\bibfield  {title} {\bibinfo {title} {Understanding bioluminescence in dinoflagellates—how far have we come?},\ }\href {https://doi.org/10.3390/microorganisms1010003} {\bibfield  {journal} {\bibinfo  {journal} {Microorganisms}\ }\textbf {\bibinfo {volume} {1}},\ \bibinfo {pages} {3} (\bibinfo {year} {2013})}\BibitemShut {NoStop}%
\bibitem [{\citenamefont {Jin}\ \emph {et~al.}(2013)\citenamefont {Jin}, \citenamefont {Klima}, \citenamefont {Deane}, \citenamefont {Stokes},\ and\ \citenamefont {Latz}}]{Latz-stretch}%
  \BibitemOpen
  \bibfield  {author} {\bibinfo {author} {\bibfnamefont {K.}~\bibnamefont {Jin}}, \bibinfo {author} {\bibfnamefont {J.~C.}\ \bibnamefont {Klima}}, \bibinfo {author} {\bibfnamefont {G.}~\bibnamefont {Deane}}, \bibinfo {author} {\bibfnamefont {M.~D.}\ \bibnamefont {Stokes}},\ and\ \bibinfo {author} {\bibfnamefont {M.~I.}\ \bibnamefont {Latz}},\ }\bibfield  {title} {\bibinfo {title} {Pharmacological investigation of the bioluminescence signaling pathway of the dinoflagellate lingulodinium polyedrum: evidence for the role of stretch-activated ion channels},\ }\href {https://doi.org/https://doi.org/10.1111/jpy.12084} {\bibfield  {journal} {\bibinfo  {journal} {Journal of Phycology}\ }\textbf {\bibinfo {volume} {49}},\ \bibinfo {pages} {733} (\bibinfo {year} {2013})}\BibitemShut {NoStop}%
\bibitem [{\citenamefont {{von Dassow}}\ and\ \citenamefont {Latz}(2002)}]{Latz-Ca}%
  \BibitemOpen
  \bibfield  {author} {\bibinfo {author} {\bibfnamefont {P.}~\bibnamefont {{von Dassow}}}\ and\ \bibinfo {author} {\bibfnamefont {M.~I.}\ \bibnamefont {Latz}},\ }\bibfield  {title} {\bibinfo {title} {The role of ca2+ in stimulated bioluminescence of the dinoflagellate lingulodinium polyedrum},\ }\href {https://doi.org/10.1242/jeb.205.19.2971} {\bibfield  {journal} {\bibinfo  {journal} {J. Exp. Biol.}\ }\textbf {\bibinfo {volume} {205}},\ \bibinfo {pages} {2971} (\bibinfo {year} {2002})}\BibitemShut {NoStop}%
\bibitem [{\citenamefont {Blaser}\ \emph {et~al.}(2002)\citenamefont {Blaser}, \citenamefont {Kurisu}, \citenamefont {Satoh},\ and\ \citenamefont {Mino}}]{Blaser2002-oscillating}%
  \BibitemOpen
  \bibfield  {author} {\bibinfo {author} {\bibfnamefont {S.}~\bibnamefont {Blaser}}, \bibinfo {author} {\bibfnamefont {F.}~\bibnamefont {Kurisu}}, \bibinfo {author} {\bibfnamefont {H.}~\bibnamefont {Satoh}},\ and\ \bibinfo {author} {\bibfnamefont {T.}~\bibnamefont {Mino}},\ }\bibfield  {title} {\bibinfo {title} {Hydromechanical stimulation of bioluminescent plankton},\ }\href {https://doi.org/https://doi.org/10.1002/bio.696} {\bibfield  {journal} {\bibinfo  {journal} {Luminescence}\ }\textbf {\bibinfo {volume} {17}},\ \bibinfo {pages} {370} (\bibinfo {year} {2002})}\BibitemShut {NoStop}%
\bibitem [{\citenamefont {Latz}\ \emph {et~al.}(2004{\natexlab{a}})\citenamefont {Latz}, \citenamefont {Juhl}, \citenamefont {Ahmed}, \citenamefont {Elghobashi},\ and\ \citenamefont {Rohr}}]{Latz04-converging}%
  \BibitemOpen
  \bibfield  {author} {\bibinfo {author} {\bibfnamefont {M.~I.}\ \bibnamefont {Latz}}, \bibinfo {author} {\bibfnamefont {A.~R.}\ \bibnamefont {Juhl}}, \bibinfo {author} {\bibfnamefont {A.~M.}\ \bibnamefont {Ahmed}}, \bibinfo {author} {\bibfnamefont {S.~E.}\ \bibnamefont {Elghobashi}},\ and\ \bibinfo {author} {\bibfnamefont {J.}~\bibnamefont {Rohr}},\ }\bibfield  {title} {\bibinfo {title} {Hydrodynamic stimulation of dinoflagellate bioluminescence: a computational and experimental study},\ }\href {https://doi.org/10.1242/jeb.00973} {\bibfield  {journal} {\bibinfo  {journal} {Journal of Experimental Biology}\ }\textbf {\bibinfo {volume} {207}},\ \bibinfo {pages} {1941} (\bibinfo {year} {2004}{\natexlab{a}})}\BibitemShut {NoStop}%
\bibitem [{\citenamefont {Latz}\ \emph {et~al.}(1994)\citenamefont {Latz}, \citenamefont {Case},\ and\ \citenamefont {Gran}}]{Latz94-Couette}%
  \BibitemOpen
  \bibfield  {author} {\bibinfo {author} {\bibfnamefont {M.~I.}\ \bibnamefont {Latz}}, \bibinfo {author} {\bibfnamefont {J.~F.}\ \bibnamefont {Case}},\ and\ \bibinfo {author} {\bibfnamefont {R.~L.}\ \bibnamefont {Gran}},\ }\bibfield  {title} {\bibinfo {title} {Excitation of bioluminescence by laminar fluid shear associated with simple couette flow},\ }\href {https://doi.org/https://doi.org/10.4319/lo.1994.39.6.1424} {\bibfield  {journal} {\bibinfo  {journal} {Limnology and Oceanography}\ }\textbf {\bibinfo {volume} {39}},\ \bibinfo {pages} {1424} (\bibinfo {year} {1994})}\BibitemShut {NoStop}%
\bibitem [{\citenamefont {Maldonado}\ and\ \citenamefont {Latz}(2007)}]{Maldonado07-shear}%
  \BibitemOpen
  \bibfield  {author} {\bibinfo {author} {\bibfnamefont {E.~M.}\ \bibnamefont {Maldonado}}\ and\ \bibinfo {author} {\bibfnamefont {M.~I.}\ \bibnamefont {Latz}},\ }\bibfield  {title} {\bibinfo {title} {Shear-stress dependence of dinoflagellate bioluminescence},\ }\href {https://doi.org/10.2307/25066606} {\bibfield  {journal} {\bibinfo  {journal} {The Biological Bulletin}\ }\textbf {\bibinfo {volume} {212}},\ \bibinfo {pages} {242} (\bibinfo {year} {2007})}\BibitemShut {NoStop}%
\bibitem [{\citenamefont {Latz}\ and\ \citenamefont {Rohr}(1999)}]{Latz1999-pipe}%
  \BibitemOpen
  \bibfield  {author} {\bibinfo {author} {\bibfnamefont {M.~I.}\ \bibnamefont {Latz}}\ and\ \bibinfo {author} {\bibfnamefont {J.}~\bibnamefont {Rohr}},\ }\bibfield  {title} {\bibinfo {title} {Luminescent response of the red tide dinoflagellate lingulodinium polyedrum to laminar and turbulent flow},\ }\href {https://doi.org/https://doi.org/10.4319/lo.1999.44.6.1423} {\bibfield  {journal} {\bibinfo  {journal} {Limnology and Oceanography}\ }\textbf {\bibinfo {volume} {44}},\ \bibinfo {pages} {1423} (\bibinfo {year} {1999})}\BibitemShut {NoStop}%
\bibitem [{\citenamefont {Rohr}\ \emph {et~al.}(2002)\citenamefont {Rohr}, \citenamefont {Hyman}, \citenamefont {Fallon},\ and\ \citenamefont {Latz}}]{Rohr2002-threshold-eps}%
  \BibitemOpen
  \bibfield  {author} {\bibinfo {author} {\bibfnamefont {J.}~\bibnamefont {Rohr}}, \bibinfo {author} {\bibfnamefont {M.}~\bibnamefont {Hyman}}, \bibinfo {author} {\bibfnamefont {S.}~\bibnamefont {Fallon}},\ and\ \bibinfo {author} {\bibfnamefont {M.~I.}\ \bibnamefont {Latz}},\ }\bibfield  {title} {\bibinfo {title} {Bioluminescence flow visualization in the ocean: an initial strategy based on laboratory experiments},\ }\href {https://doi.org/https://doi.org/10.1016/S0967-0637(02)00116-4} {\bibfield  {journal} {\bibinfo  {journal} {Deep Sea Research Part I: Oceanographic Research Papers}\ }\textbf {\bibinfo {volume} {49}},\ \bibinfo {pages} {2009} (\bibinfo {year} {2002})}\BibitemShut {NoStop}%
\bibitem [{\citenamefont {Rohr}\ \emph {et~al.}(1995)\citenamefont {Rohr}, \citenamefont {Latz}, \citenamefont {Hendricks},\ and\ \citenamefont {Nauen}}]{Latz95-visual2}%
  \BibitemOpen
  \bibfield  {author} {\bibinfo {author} {\bibfnamefont {J.}~\bibnamefont {Rohr}}, \bibinfo {author} {\bibfnamefont {M.~I.}\ \bibnamefont {Latz}}, \bibinfo {author} {\bibfnamefont {E.}~\bibnamefont {Hendricks}},\ and\ \bibinfo {author} {\bibfnamefont {J.~C.}\ \bibnamefont {Nauen}},\ }\bibfield  {title} {\bibinfo {title} {A novel flow visualization technique using bioluminescent marine plankton ii. field studies},\ }\href {https://doi.org/10.1109/48.376679} {\bibfield  {journal} {\bibinfo  {journal} {IEEE Journal of Oceanic Engineering}\ }\textbf {\bibinfo {volume} {20}},\ \bibinfo {pages} {147} (\bibinfo {year} {1995})}\BibitemShut {NoStop}%
\bibitem [{\citenamefont {Stokes}\ \emph {et~al.}(2004)\citenamefont {Stokes}, \citenamefont {Deane}, \citenamefont {Latz},\ and\ \citenamefont {Rohr}}]{Stokes04-waves}%
  \BibitemOpen
  \bibfield  {author} {\bibinfo {author} {\bibfnamefont {M.~D.}\ \bibnamefont {Stokes}}, \bibinfo {author} {\bibfnamefont {G.~B.}\ \bibnamefont {Deane}}, \bibinfo {author} {\bibfnamefont {M.~I.}\ \bibnamefont {Latz}},\ and\ \bibinfo {author} {\bibfnamefont {J.}~\bibnamefont {Rohr}},\ }\bibfield  {title} {\bibinfo {title} {Bioluminescence imaging of wave-induced turbulence},\ }\href {https://doi.org/https://doi.org/10.1029/2003JC001871} {\bibfield  {journal} {\bibinfo  {journal} {Journal of Geophysical Research: Oceans}\ }\textbf {\bibinfo {volume} {109}} (\bibinfo {year} {2004})}\BibitemShut {NoStop}%
\bibitem [{\citenamefont {von Dassow}\ \emph {et~al.}(2005)\citenamefont {von Dassow}, \citenamefont {Bearon},\ and\ \citenamefont {Latz}}]{Latz2005-transient}%
  \BibitemOpen
  \bibfield  {author} {\bibinfo {author} {\bibfnamefont {P.}~\bibnamefont {von Dassow}}, \bibinfo {author} {\bibfnamefont {R.~N.}\ \bibnamefont {Bearon}},\ and\ \bibinfo {author} {\bibfnamefont {M.~I.}\ \bibnamefont {Latz}},\ }\bibfield  {title} {\bibinfo {title} {Bioluminescent response of the dinoflagellate lingulodinium polyedrum to developing flow: Tuning of sensitivity and the role of desensitization in controlling a defensive behavior of a planktonic cell},\ }\href {https://doi.org/https://doi.org/10.4319/lo.2005.50.2.0607} {\bibfield  {journal} {\bibinfo  {journal} {Limnology and Oceanography}\ }\textbf {\bibinfo {volume} {50}},\ \bibinfo {pages} {607} (\bibinfo {year} {2005})}\BibitemShut {NoStop}%
\bibitem [{\citenamefont {Deane}\ and\ \citenamefont {Stokes}(2005)}]{Stokes2005-model}%
  \BibitemOpen
  \bibfield  {author} {\bibinfo {author} {\bibfnamefont {G.~B.}\ \bibnamefont {Deane}}\ and\ \bibinfo {author} {\bibfnamefont {M.~D.}\ \bibnamefont {Stokes}},\ }\bibfield  {title} {\bibinfo {title} {A quantitative model for flow-induced bioluminescence in dinoflagellates},\ }\href {https://doi.org/https://doi.org/10.1016/j.jtbi.2005.04.002} {\bibfield  {journal} {\bibinfo  {journal} {Journal of Theoretical Biology}\ }\textbf {\bibinfo {volume} {237}},\ \bibinfo {pages} {147} (\bibinfo {year} {2005})}\BibitemShut {NoStop}%
\bibitem [{\citenamefont {Jalaal}\ \emph {et~al.}(2020)\citenamefont {Jalaal}, \citenamefont {Schramma}, \citenamefont {Dode}, \citenamefont {de~Maleprade}, \citenamefont {Raufaste},\ and\ \citenamefont {Goldstein}}]{jalaal2020}%
  \BibitemOpen
  \bibfield  {author} {\bibinfo {author} {\bibfnamefont {M.}~\bibnamefont {Jalaal}}, \bibinfo {author} {\bibfnamefont {N.}~\bibnamefont {Schramma}}, \bibinfo {author} {\bibfnamefont {A.}~\bibnamefont {Dode}}, \bibinfo {author} {\bibfnamefont {H.}~\bibnamefont {de~Maleprade}}, \bibinfo {author} {\bibfnamefont {C.}~\bibnamefont {Raufaste}},\ and\ \bibinfo {author} {\bibfnamefont {R.~E.}\ \bibnamefont {Goldstein}},\ }\bibfield  {title} {\bibinfo {title} {Stress-induced dinoflagellate bioluminescence at the single cell level},\ }\href {https://doi.org/10.1103/PhysRevLett.125.028102} {\bibfield  {journal} {\bibinfo  {journal} {Phys. Rev. Lett.}\ }\textbf {\bibinfo {volume} {125}},\ \bibinfo {pages} {028102} (\bibinfo {year} {2020})}\BibitemShut {NoStop}%
\bibitem [{\citenamefont {Balkovsky}\ \emph {et~al.}(2000)\citenamefont {Balkovsky}, \citenamefont {Fouxon},\ and\ \citenamefont {Lebedev}}]{bfl00}%
  \BibitemOpen
  \bibfield  {author} {\bibinfo {author} {\bibfnamefont {E.}~\bibnamefont {Balkovsky}}, \bibinfo {author} {\bibfnamefont {A.}~\bibnamefont {Fouxon}},\ and\ \bibinfo {author} {\bibfnamefont {V.}~\bibnamefont {Lebedev}},\ }\bibfield  {title} {\bibinfo {title} {Turbulent dynamics of polymer solutions},\ }\href@noop {} {\bibfield  {journal} {\bibinfo  {journal} {Phys. Rev. Lett.}\ }\textbf {\bibinfo {volume} {84}},\ \bibinfo {pages} {4765} (\bibinfo {year} {2000})}\BibitemShut {NoStop}%
\bibitem [{\citenamefont {Balkovsky}\ \emph {et~al.}(2001)\citenamefont {Balkovsky}, \citenamefont {Fouxon},\ and\ \citenamefont {Lebedev}}]{bfl01}%
  \BibitemOpen
  \bibfield  {author} {\bibinfo {author} {\bibfnamefont {E.}~\bibnamefont {Balkovsky}}, \bibinfo {author} {\bibfnamefont {A.}~\bibnamefont {Fouxon}},\ and\ \bibinfo {author} {\bibfnamefont {V.}~\bibnamefont {Lebedev}},\ }\bibfield  {title} {\bibinfo {title} {Turbulence of polymer solutions},\ }\href@noop {} {\bibfield  {journal} {\bibinfo  {journal} {Phys. Rev. E}\ }\textbf {\bibinfo {volume} {64}},\ \bibinfo {pages} {056301} (\bibinfo {year} {2001})}\BibitemShut {NoStop}%
\bibitem [{\citenamefont {Chertkov}(2000)}]{c00}%
  \BibitemOpen
  \bibfield  {author} {\bibinfo {author} {\bibfnamefont {M.}~\bibnamefont {Chertkov}},\ }\bibfield  {title} {\bibinfo {title} {Polymer stretching by turbulence},\ }\href@noop {} {\bibfield  {journal} {\bibinfo  {journal} {Phys. Rev. Lett.}\ }\textbf {\bibinfo {volume} {84}},\ \bibinfo {pages} {4761} (\bibinfo {year} {2000})}\BibitemShut {NoStop}%
\bibitem [{\citenamefont {Ishihara}\ \emph {et~al.}(2007)\citenamefont {Ishihara}, \citenamefont {Kaneda}, \citenamefont {Yokokawa}, \citenamefont {Itakura},\ and\ \citenamefont {Uno}}]{Ishihara2007}%
  \BibitemOpen
  \bibfield  {author} {\bibinfo {author} {\bibfnamefont {T.}~\bibnamefont {Ishihara}}, \bibinfo {author} {\bibfnamefont {Y.}~\bibnamefont {Kaneda}}, \bibinfo {author} {\bibfnamefont {M.}~\bibnamefont {Yokokawa}}, \bibinfo {author} {\bibfnamefont {K.}~\bibnamefont {Itakura}},\ and\ \bibinfo {author} {\bibfnamefont {A.}~\bibnamefont {Uno}},\ }\bibfield  {title} {\bibinfo {title} {Small-scale statistics in high-resolution direct numerical simulation of turbulence: Reynolds number dependence of one-point velocity gradient statistics},\ }\href@noop {} {\bibfield  {journal} {\bibinfo  {journal} {J. Fluid Mech.}\ }\textbf {\bibinfo {volume} {592}},\ \bibinfo {pages} {335–366} (\bibinfo {year} {2007})}\BibitemShut {NoStop}%
\bibitem [{\citenamefont {Buaria}\ \emph {et~al.}(2019)\citenamefont {Buaria}, \citenamefont {Pumir}, \citenamefont {Bodenschatz},\ and\ \citenamefont {Yeung}}]{Buaria2019}%
  \BibitemOpen
  \bibfield  {author} {\bibinfo {author} {\bibfnamefont {D.}~\bibnamefont {Buaria}}, \bibinfo {author} {\bibfnamefont {A.}~\bibnamefont {Pumir}}, \bibinfo {author} {\bibfnamefont {E.}~\bibnamefont {Bodenschatz}},\ and\ \bibinfo {author} {\bibfnamefont {P.~K.}\ \bibnamefont {Yeung}},\ }\bibfield  {title} {\bibinfo {title} {Extreme velocity gradients in turbulent flows},\ }\href@noop {} {\bibfield  {journal} {\bibinfo  {journal} {New J. Phys.}\ }\textbf {\bibinfo {volume} {21}},\ \bibinfo {pages} {043004} (\bibinfo {year} {2019})}\BibitemShut {NoStop}%
\bibitem [{\citenamefont {Buaria}\ and\ \citenamefont {Pumir}(2022)}]{Buaria2022}%
  \BibitemOpen
  \bibfield  {author} {\bibinfo {author} {\bibfnamefont {D.}~\bibnamefont {Buaria}}\ and\ \bibinfo {author} {\bibfnamefont {A.}~\bibnamefont {Pumir}},\ }\bibfield  {title} {\bibinfo {title} {Vorticity-strain rate dynamics and the smallest scales of turbulence},\ }\href {https://doi.org/10.1103/PhysRevLett.128.094501} {\bibfield  {journal} {\bibinfo  {journal} {Phys. Rev. Lett.}\ }\textbf {\bibinfo {volume} {128}},\ \bibinfo {pages} {094501} (\bibinfo {year} {2022})}\BibitemShut {NoStop}%
\bibitem [{\citenamefont {Franks}\ \emph {et~al.}(2022)\citenamefont {Franks}, \citenamefont {Inman}, \citenamefont {MacKinnon}, \citenamefont {Alford},\ and\ \citenamefont {Waterhouse}}]{Franks22-eps}%
  \BibitemOpen
  \bibfield  {author} {\bibinfo {author} {\bibfnamefont {P.~J.~S.}\ \bibnamefont {Franks}}, \bibinfo {author} {\bibfnamefont {B.~G.}\ \bibnamefont {Inman}}, \bibinfo {author} {\bibfnamefont {J.~A.}\ \bibnamefont {MacKinnon}}, \bibinfo {author} {\bibfnamefont {M.~H.}\ \bibnamefont {Alford}},\ and\ \bibinfo {author} {\bibfnamefont {A.~F.}\ \bibnamefont {Waterhouse}},\ }\bibfield  {title} {\bibinfo {title} {Oceanic turbulence from a planktonic perspective},\ }\href {https://doi.org/https://doi.org/10.1002/lno.11996} {\bibfield  {journal} {\bibinfo  {journal} {Limnology and Oceanography}\ }\textbf {\bibinfo {volume} {67}},\ \bibinfo {pages} {348} (\bibinfo {year} {2022})}\BibitemShut {NoStop}%
\bibitem [{\citenamefont {Franks}\ and\ \citenamefont {Inman}(2024)}]{Franks24-gradv}%
  \BibitemOpen
  \bibfield  {author} {\bibinfo {author} {\bibfnamefont {P.~J.~S.}\ \bibnamefont {Franks}}\ and\ \bibinfo {author} {\bibfnamefont {B.~G.}\ \bibnamefont {Inman}},\ }\bibfield  {title} {\bibinfo {title} {Shortcomings of the dissipation rate for understanding the turbulent environment of plankton—and a potential solution},\ }\href {https://doi.org/https://doi.org/10.1002/lno.12501} {\bibfield  {journal} {\bibinfo  {journal} {Limnology and Oceanography}\ }\textbf {\bibinfo {volume} {69}},\ \bibinfo {pages} {S88} (\bibinfo {year} {2024})}\BibitemShut {NoStop}%
\bibitem [{\citenamefont {Latz}\ \emph {et~al.}(2004{\natexlab{b}})\citenamefont {Latz}, \citenamefont {Nauen},\ and\ \citenamefont {Rohr}}]{Latz04-4species-threshold}%
  \BibitemOpen
  \bibfield  {author} {\bibinfo {author} {\bibfnamefont {M.~I.}\ \bibnamefont {Latz}}, \bibinfo {author} {\bibfnamefont {J.~C.}\ \bibnamefont {Nauen}},\ and\ \bibinfo {author} {\bibfnamefont {J.}~\bibnamefont {Rohr}},\ }\bibfield  {title} {\bibinfo {title} {Bioluminescence response of four species of dinoflagellates to fully developed pipe flow},\ }\href {https://doi.org/10.1093/plankt/fbh141} {\bibfield  {journal} {\bibinfo  {journal} {Journal of Plankton Research}\ }\textbf {\bibinfo {volume} {26}},\ \bibinfo {pages} {1529} (\bibinfo {year} {2004}{\natexlab{b}})}\BibitemShut {NoStop}%
\bibitem [{sup()}]{supplement}%
  \BibitemOpen
  \href@noop {} {}\bibinfo {note} {See Supplementary Material at \url{https://bighome.iitb.ac.in/index.php/s/SFqx6MTagxonP8i} for additional results.}\BibitemShut {Stop}%
\bibitem [{\citenamefont {James}\ and\ \citenamefont {Ray}(2017)}]{jr17}%
  \BibitemOpen
  \bibfield  {author} {\bibinfo {author} {\bibfnamefont {M.}~\bibnamefont {James}}\ and\ \bibinfo {author} {\bibfnamefont {S.~S.}\ \bibnamefont {Ray}},\ }\bibfield  {title} {\bibinfo {title} {Enhanced droplet collision rates and impact velocities in turbulent flows: The effect of poly-dispersity and transient phases},\ }\href@noop {} {\bibfield  {journal} {\bibinfo  {journal} {Sci. Reports}\ }\textbf {\bibinfo {volume} {7}},\ \bibinfo {pages} {12231} (\bibinfo {year} {2017})}\BibitemShut {NoStop}%
\bibitem [{\citenamefont {Picardo}\ \emph {et~al.}(2023)\citenamefont {Picardo}, \citenamefont {Plan},\ and\ \citenamefont {Vincenzi}}]{ppv23}%
  \BibitemOpen
  \bibfield  {author} {\bibinfo {author} {\bibfnamefont {J.~R.}\ \bibnamefont {Picardo}}, \bibinfo {author} {\bibfnamefont {E.~L.~C. V.~M.}\ \bibnamefont {Plan}},\ and\ \bibinfo {author} {\bibfnamefont {D.}~\bibnamefont {Vincenzi}},\ }\bibfield  {title} {\bibinfo {title} {Polymers in turbulence: stretching statistics and the role of extreme strain rate fluctuations},\ }\href@noop {} {\bibfield  {journal} {\bibinfo  {journal} {J. Fluid Mech.}\ }\textbf {\bibinfo {volume} {969}},\ \bibinfo {pages} {A24} (\bibinfo {year} {2023})}\BibitemShut {NoStop}%
\bibitem [{\citenamefont {Ganesh}\ \emph {et~al.}(2025)\citenamefont {Ganesh}, \citenamefont {Vincenzi}, \citenamefont {Prabhakar},\ and\ \citenamefont {Picardo}}]{gvpp25}%
  \BibitemOpen
  \bibfield  {author} {\bibinfo {author} {\bibfnamefont {A.}~\bibnamefont {Ganesh}}, \bibinfo {author} {\bibfnamefont {D.}~\bibnamefont {Vincenzi}}, \bibinfo {author} {\bibfnamefont {R.}~\bibnamefont {Prabhakar}},\ and\ \bibinfo {author} {\bibfnamefont {J.~R.}\ \bibnamefont {Picardo}},\ }\bibfield  {title} {\bibinfo {title} {How hydrodynamic interactions alter polymer stretching in turbulence}} (\bibinfo {year} {2025}),\ \bibinfo {note} {\texttt{arXiv}}\BibitemShut {NoStop}%
\bibitem [{\citenamefont {Dieci}\ \emph {et~al.}(1997)\citenamefont {Dieci}, \citenamefont {Russell},\ and\ \citenamefont {Vleck}}]{drv97}%
  \BibitemOpen
  \bibfield  {author} {\bibinfo {author} {\bibfnamefont {L.}~\bibnamefont {Dieci}}, \bibinfo {author} {\bibfnamefont {R.~D.}\ \bibnamefont {Russell}},\ and\ \bibinfo {author} {\bibfnamefont {E.~S.~V.}\ \bibnamefont {Vleck}},\ }\bibfield  {title} {\bibinfo {title} {On the computation of {L}yapunov exponents for continuous dynamical systems},\ }\href@noop {} {\bibfield  {journal} {\bibinfo  {journal} {SIAM J. Numer. Anal.}\ }\textbf {\bibinfo {volume} {34}},\ \bibinfo {pages} {402} (\bibinfo {year} {1997})}\BibitemShut {NoStop}%
\bibitem [{\citenamefont {Brunk}\ \emph {et~al.}(1998)\citenamefont {Brunk}, \citenamefont {Koch},\ and\ \citenamefont {Lion}}]{bkl98}%
  \BibitemOpen
  \bibfield  {author} {\bibinfo {author} {\bibfnamefont {B.~K.}\ \bibnamefont {Brunk}}, \bibinfo {author} {\bibfnamefont {D.~L.}\ \bibnamefont {Koch}},\ and\ \bibinfo {author} {\bibfnamefont {L.~W.}\ \bibnamefont {Lion}},\ }\bibfield  {title} {\bibinfo {title} {Turbulent coagulation of colloidal particles},\ }\href@noop {} {\bibfield  {journal} {\bibinfo  {journal} {J. Fluid Mech.}\ }\textbf {\bibinfo {volume} {364}},\ \bibinfo {pages} {81} (\bibinfo {year} {1998})}\BibitemShut {NoStop}%
\bibitem [{\citenamefont {Frisch}(1995)}]{f95}%
  \BibitemOpen
  \bibfield  {author} {\bibinfo {author} {\bibfnamefont {U.}~\bibnamefont {Frisch}},\ }\href@noop {} {\emph {\bibinfo {title} {Turbulence: The Legacy of A. N. Kolmogorov}}}\ (\bibinfo  {publisher} {Cambridge University Press},\ \bibinfo {address} {Cambridge, England},\ \bibinfo {year} {1995})\BibitemShut {NoStop}%
\bibitem [{\citenamefont {Bec}\ \emph {et~al.}(2006)\citenamefont {Bec}, \citenamefont {Biferale}, \citenamefont {Boffetta}, \citenamefont {Cencini}, \citenamefont {Musacchio},\ and\ \citenamefont {Toschi}}]{bbbcmt06}%
  \BibitemOpen
  \bibfield  {author} {\bibinfo {author} {\bibfnamefont {J.}~\bibnamefont {Bec}}, \bibinfo {author} {\bibfnamefont {L.}~\bibnamefont {Biferale}}, \bibinfo {author} {\bibfnamefont {G.}~\bibnamefont {Boffetta}}, \bibinfo {author} {\bibfnamefont {M.}~\bibnamefont {Cencini}}, \bibinfo {author} {\bibfnamefont {S.}~\bibnamefont {Musacchio}},\ and\ \bibinfo {author} {\bibfnamefont {F.}~\bibnamefont {Toschi}},\ }\bibfield  {title} {\bibinfo {title} {{Lyapunov} exponents of heavy particles in turbulence},\ }\href@noop {} {\bibfield  {journal} {\bibinfo  {journal} {Phys. Fluids}\ }\textbf {\bibinfo {volume} {18}},\ \bibinfo {pages} {091702} (\bibinfo {year} {2006})}\BibitemShut {NoStop}%
\bibitem [{\citenamefont {Pujara}\ \emph {et~al.}(2018)\citenamefont {Pujara}, \citenamefont {Koehl},\ and\ \citenamefont {Variano}}]{Pujara2018}%
  \BibitemOpen
  \bibfield  {author} {\bibinfo {author} {\bibfnamefont {N.}~\bibnamefont {Pujara}}, \bibinfo {author} {\bibfnamefont {M.~A.~R.}\ \bibnamefont {Koehl}},\ and\ \bibinfo {author} {\bibfnamefont {E.~A.}\ \bibnamefont {Variano}},\ }\bibfield  {title} {\bibinfo {title} {Rotations and accumulation of ellipsoidal microswimmers in isotropic turbulence},\ }\href {https://doi.org/10.1017/jfm.2017.912} {\bibfield  {journal} {\bibinfo  {journal} {J. Fluid Mech.}\ }\textbf {\bibinfo {volume} {838}},\ \bibinfo {pages} {356–368} (\bibinfo {year} {2018})}\BibitemShut {NoStop}%
\bibitem [{\citenamefont {Picardo}\ \emph {et~al.}(2020)\citenamefont {Picardo}, \citenamefont {Singh}, \citenamefont {Ray},\ and\ \citenamefont {Vincenzi}}]{Picardo2020}%
  \BibitemOpen
  \bibfield  {author} {\bibinfo {author} {\bibfnamefont {J.~R.}\ \bibnamefont {Picardo}}, \bibinfo {author} {\bibfnamefont {R.}~\bibnamefont {Singh}}, \bibinfo {author} {\bibfnamefont {S.~S.}\ \bibnamefont {Ray}},\ and\ \bibinfo {author} {\bibfnamefont {D.}~\bibnamefont {Vincenzi}},\ }\bibfield  {title} {\bibinfo {title} {Dynamics of a long chain in turbulent flows: impact of vortices},\ }\href {https://doi.org/10.1098/rsta.2019.0405} {\bibfield  {journal} {\bibinfo  {journal} {Philos. Trans. R. Soc. A}\ }\textbf {\bibinfo {volume} {378}},\ \bibinfo {pages} {20190405} (\bibinfo {year} {2020})}\BibitemShut {NoStop}%
\bibitem [{\citenamefont {Breier}\ \emph {et~al.}(2018)\citenamefont {Breier}, \citenamefont {Lalescu}, \citenamefont {Waas}, \citenamefont {Wilczek},\ and\ \citenamefont {Mazza}}]{Wilczek2018}%
  \BibitemOpen
  \bibfield  {author} {\bibinfo {author} {\bibfnamefont {R.~E.}\ \bibnamefont {Breier}}, \bibinfo {author} {\bibfnamefont {C.~C.}\ \bibnamefont {Lalescu}}, \bibinfo {author} {\bibfnamefont {D.}~\bibnamefont {Waas}}, \bibinfo {author} {\bibfnamefont {M.}~\bibnamefont {Wilczek}},\ and\ \bibinfo {author} {\bibfnamefont {M.~G.}\ \bibnamefont {Mazza}},\ }\bibfield  {title} {\bibinfo {title} {Emergence of phytoplankton patchiness at small scales in mild turbulence},\ }\href {https://doi.org/10.1073/pnas.1808711115} {\bibfield  {journal} {\bibinfo  {journal} {PNAS}\ }\textbf {\bibinfo {volume} {115}},\ \bibinfo {pages} {12112} (\bibinfo {year} {2018})}\BibitemShut {NoStop}%
\bibitem [{\citenamefont {Gao}\ \emph {et~al.}(2011)\citenamefont {Gao}, \citenamefont {Hu},\ and\ \citenamefont {Castañeda}}]{Gao2011}%
  \BibitemOpen
  \bibfield  {author} {\bibinfo {author} {\bibfnamefont {T.}~\bibnamefont {Gao}}, \bibinfo {author} {\bibfnamefont {H.~H.}\ \bibnamefont {Hu}},\ and\ \bibinfo {author} {\bibfnamefont {P.~P.}\ \bibnamefont {Castañeda}},\ }\bibfield  {title} {\bibinfo {title} {Rheology of a suspension of elastic particles in a viscous shear flow},\ }\href {https://doi.org/10.1017/jfm.2011.347} {\bibfield  {journal} {\bibinfo  {journal} {J. Fluid Mech.}\ }\textbf {\bibinfo {volume} {687}},\ \bibinfo {pages} {209–237} (\bibinfo {year} {2011})}\BibitemShut {NoStop}%
\bibitem [{\citenamefont {Sanagavarapu}\ \emph {et~al.}(2022)\citenamefont {Sanagavarapu}, \citenamefont {Subramanian},\ and\ \citenamefont {Nott}}]{Nott2022}%
  \BibitemOpen
  \bibfield  {author} {\bibinfo {author} {\bibfnamefont {P.~K.}\ \bibnamefont {Sanagavarapu}}, \bibinfo {author} {\bibfnamefont {G.}~\bibnamefont {Subramanian}},\ and\ \bibinfo {author} {\bibfnamefont {P.~R.}\ \bibnamefont {Nott}},\ }\bibfield  {title} {\bibinfo {title} {Shape dynamics and rheology of dilute suspensions of elastic and viscoelastic particles},\ }\href {https://doi.org/10.1017/jfm.2022.704} {\bibfield  {journal} {\bibinfo  {journal} {J. Fluid Mech.}\ }\textbf {\bibinfo {volume} {949}},\ \bibinfo {pages} {A22} (\bibinfo {year} {2022})}\BibitemShut {NoStop}%
\bibitem [{\citenamefont {Hao}(2024)}]{Hao24}%
  \BibitemOpen
  \bibfield  {author} {\bibinfo {author} {\bibfnamefont {X.}~\bibnamefont {Hao}},\ }\bibfield  {title} {\bibinfo {title} {Quantifying bioluminescent light intensity in breaking waves using numerical simulations},\ }\href {https://doi.org/https://doi.org/10.1029/2024GL110884} {\bibfield  {journal} {\bibinfo  {journal} {Geophys. Res. Lett.}\ }\textbf {\bibinfo {volume} {51}},\ \bibinfo {pages} {e2024GL110884} (\bibinfo {year} {2024})}\BibitemShut {NoStop}%
\end{thebibliography}%

\end{document}